\documentclass[preprint2,numberedappendix]{emulateapj-rtx4}
\usepackage{amsmath}
\usepackage{amssymb}

\usepackage{esint}

\usepackage{bm,color}

\def\solphys{Sol. Phys.}
\def\aap{A\&A}
\def\aa{Astron. Astrophys.}

\def\apj{Astrophys. J.}
\def\apjl{Astrophys. J. Lett.}

\def\jgr{J. Geophys. Res.}
\def\mnras{Mon. Not. R. Astron. Soc.}

\def\solphys{Solar Phys.}

\def\sp{Solar Phys.}

\begin{document}

\title{Injection of magnetic helicity in solar cycle 24 and early phase of cycle 25}%{Injection of magnetic helicity in solar cycle 24 and early phase of cycle 25}

\author{H. Q. Zhang$^1$, S. B. Yang$^{1}$ , K. M. Kuzanyan$^{1,2}$, Axel Brandenburg$^{3,4}$, H. Q. Xu$^{1}$, and D. D. Sokoloff$^{2,5}$}
\affil{
$^1$National Astronomical Observatories, Chinese Academy of Sciences, Beijing 100101, China, \\
%E-mail: hzhang@bao.ac.cn\\
$^{2}$IZMIRAN, Russian Academy of Sciences, Troitsk, Moscow 142190, Russia\\  
$^3$Nordita, KTH Royal Institute of Technology and Stockholm University, Hannes Alfv\'ens v\"ag 12, 10691\\
 Stockholm, Sweden\\
$^{4}$The Oskar Klein Centre, Department of Astronomy, Stockholm University, AlbaNova, 10691 \\
Stockholm, Sweden\\
$^5$Department of Physics, Moscow University, 119992 Moscow, Russia
}

%\submitted{\today,~ $ $Revision: 1.97 $ $}

% These dates will be filled out by the publisher
%\date{Accepted XXX. Received YYY; in original form ZZZ}

\begin{abstract}

The injection of magnetic helicity into the heliosphere during solar cycle 24 and the early phase of cycle 25
has been calculated based on the analysis of a series of synoptic magnetic charts.
During the cycle, the injected magnetic helicity is found to be mainly contributed by the magnetic field in active regions.
According to Hale's law, the polarities of active regions statistically reverse between solar cycles 24 and 25.
We suggest that the dominant source of injected magnetic helicity likely arises
from the relatively strong magnetic fields of the leading polarity of active regions.
This occurs as part of the magnetic field that migrates to high latitudes and the polar regions of the Sun
due to the effect of meridional circulation inferred from a series of HMI/SDO magnetic synoptic charts.
Significant fluctuations of the injected magnetic helicity from the subsurface layers may reflect the complex processes
of how the twist from the convection zone ejects magnetic fields through a series of active regions
on different temporal and spatial scales at the solar surface.

\end{abstract}


\email{hzhang@bao.ac.cn}
%%%%%%%%%%%%%%%%% BODY OF PAPER %%%%%%%%%%%%%%%%%%

\section{Introduction}

Helicity is an important quantity that reflects the topology of the magnetic field, such as linkage, twist, and writhe of the field lines \citep{Woltjer58a,Woltjer58b,Taylor86}. It is an integral measure of the topological properties of the field in a closed volume $V$:
\begin{equation}
\label{ }
H_f =\int{\bf F}\cdot\nabla \times {\bf F} \, dV,
\end{equation}
where $\bf F$ can be the magnetic vector potential {\bf A}, the magnetic field {\bf B}, or the velocity field {\bf V}, which is assumed to be confined to the volume $V$.

In the solar atmosphere, the magnetic and velocity fields are observable quantities measured by vector magnetographs, while the magnetic potential {\bf A} and the electric current density {\bf J} $\left(=\nabla\times{\bf B}/\mu_0\right)$ are derived quantities under some assumptions \citep[cf.][]{Zhang00,Zhang23}.
The computation of the magnetic helicity in the Sun requires knowledge of the full vector field in a 3D volume,
but observations of the magnetic fields are usually taken in a shallow layer of the solar atmosphere (typically, in the photosphere). Due to the limitation of the observations of vector magnetic fields in a single layer of the solar atmosphere, constructing the real distribution of all components of the current density in the solar photosphere remains challenging \citep{Xu15}. This implies that we still cannot get all components of the current density, not even in the lower solar photosphere. A similar constraint applies to the magnetic potential {\bf A}. Consequently, the completeness of helicity calculations based on vector magnetogram observations is inherently compromised.

Nevertheless, the magnetic helicity density $h_{\rm m}={\bf A}\cdot\nabla\times {\bf A}$,
current helicity density $h_c={\bf B}\cdot\nabla\times {\bf B}$, and velocity helicity density $h_v={\bf V}\cdot\nabla\times {\bf V}$ are still important detectable quantities. For example, one can calculate a part of the current helicity density $h_c=3({\bf B})_z\cdot(\nabla\times {\bf B})_z$ inferred from observed photospheric vector magnetograms under the isotropic assumption \citep{Xu15} and the injection rate of magnetic helicity from the solar subsurface regions \citep{cha01} based on the photospheric magnetograms. This implies that the analysis of helicities involves observable quantities in the solar atmosphere only. In addition, it is normally believed that the complex distribution of magnetic (current) helicity density in the solar active regions relates to the trigger of powerful flares and CMEs \citep[cf.][]{Bao99,liu06,Zhangy08}.

The helical topology of magnetic fields in active regions was first observed from the handedness of sunspot penumbral configurations by \cite{Hale08} and statistically with respect to hemispheres by \cite{Ding87}. The hemispheric sign rule of helicity was subsequently analyzed by \cite{Seehafer90,Pevtsov95,BZ98,Hagino05}, who found that the current helicity density and twist in solar active regions follow the hemispheric helicity rule with predominantly negative values in the northern hemisphere and positive values in the southern hemisphere. It is noted that a violation of the magnetic helicity sign rule in the beginning of each solar cycle was discovered by \cite{bao00} based on observations of a series of vector magnetograms at Huairou Solar Observing Station and a reversal of the helicity sign rule with time by \cite{Hagino05} from the observations at National Astronomical Observatory of Japan. The distribution of current helicity of active regions in solar cycles 22 and 23 was presented by \cite{zetal10}, the injection of magnetic helicity by \cite{YangSB12} and \cite{Zhang13}, and also the magnetic helicity by \cite{pipin19} for solar cycle 24 based on the vector field synoptic maps.

A large amount of observational data on solar magnetic fields provides crucial information and a foundation for theories regarding the formation of solar magnetic fields. Magnetic helicity plays a crucial role in solar magnetic field theory, and it is a key parameter connected with solar magnetic field observations. The importance of magnetic helicity was recognized early in the development of solar dynamo theory, which incorporates velocity field turbulence in the solar convection zone (i.e., the mean field dynamo stage) \citep[cf.][]{pouquet-al:1975,kleruz82,bra-sub:05}. Employing a flux transport dynamo model, \cite{Choudhuri04} explained that the hemispheric helicity could reverse sign at the beginning of each cycle, as seen in the observations of \cite{bao00}. A similar work has been proposed based on the analytical solution of the mean field dynamo model by \cite{Xu09}, who found that a reversed sign of mean hemispheric helicity can also occur in the decaying stage of a cycle.
However, the more complex distribution of magnetic helicity with the solar cycle can be found from the butterfly diagrams of mean current helicity of solar active regions in solar cycles 22 and 23 \citep{zetal10}, and it is also confirmed from the injection of global magnetic helicity from the solar surface by \cite{YangSB12} and \cite{Zhang13}. These imply that the formation of magnetic fields inside the Sun is a complex process, which has important implications for understanding the solar dynamo mechanism and predicting solar activity. 

Section~2 presents the injection of magnetic helicity during solar cycle 24 and the early phase of cycle 25 inferred by the large-scale magnetic fields. Section~3 describes the magnetic helicity and the relationship with the evolution of magnetic fields in the individual active regions,  while Section~4 encompasses discussions and conclusions.

\section{Method}

The transfer of magnetic helicity is accompanied by the dynamo process within the Sun. Corresponding  discussions have been provided by \cite{kleruz82}
and \cite{ketal95}.
The variation of the mean value of the fluctuating small-scale magnetic helicity, $h_{\rm m}=\langle{\bf a\cdot b}\rangle$, can be written in the form
\begin{equation}
\label{eq:heliflux3}
\begin{aligned}
\frac{\partial h_{\rm m}}{\partial t}=&-2\langle({\bf u}\times {\bf b})\cdot {\bf B}\rangle-2\eta\langle{\bf b}\cdot(\nabla\times {\bf b})\rangle\\
&-\langle\nabla\cdot[{\bf a}\times ({\bf u\times b})]\rangle,
\end{aligned}
\end{equation}
where {\bf b}, {\bf u}, and {\bf B} are the fluctuating magnetic field, the fluctuating velocity field, and the mean magnetic field, respectively,
and $\eta$ is the microphysical magnetic diffusivity.
The effective electromotive force is give by ${\bf E}_\mathrm{eff}\!\equiv\langle{\bf u}\times {\bf b}\rangle\!=\alpha {\bf B}\!-\!\eta_\mathrm{T}(\nabla\times {\bf B})$,
where $\eta_\mathrm{T}$ is the turbulent magnetic diffusivity
\citep{M78,Parker1979CMF,Zeldovich83,bra-sub:05}.
Even if the total magnetic helicity is conserved,
the integrals of the magnetic helicity in equation~(\ref{eq:heliflux3}) in the northern and southern hemispheres can be written as
\begin{equation}
\label{eq:heliflux}
\begin{aligned}
   \frac{d H_{\rm m}}{d t}=&\int_V2[\eta_\mathrm{T}{\bf B}\cdot(\nabla\times {\bf B})-\alpha{\bf B}^2-\eta\langle{\bf b}\cdot(\nabla\times {\bf b})\rangle]dV\\
   &-\oiint_S{\bf F}_{\rm hel}\cdot d{\bf S},
\end{aligned}
\end{equation}
where the helicity flux ${\bf F}_{\rm hel}= \bf \langle a\cdot b\rangle V-\langle(V\cdot a)b\rangle   +\langle a\times u\rangle \times B+\langle a\times(u\times b)\rangle \cdots$,
$\bf a$ is the fluctuating magnetic vector potential, $\bf V$ is the mean velocity field, and $\alpha=\alpha_v+\alpha_{\rm m}$ is the total $\alpha$ effect
\citep{kle-rog99,bra-sub:05}.
The first term in the right-hand side of equation (\ref{eq:heliflux}) relates to the generation of magnetic helicity field inside the domain $V$ (such as the Sun) and the second to the injection of helicity flux from the surface $S$ in the exterior of the solar dynamo.

Given the observation of magnetic fields, the injected magnetic helicity, which is related to the twist or linkage of magnetic fields at the surface, can be written in the form of
\citep{BergerField84,Demoulin03}
\begin{equation}
\label{eq:heliflux1}
\begin{aligned}
F_{\rm m}&=\frac{dH_{\rm m}}{dt}
=-2\oiint_S[({\bf V}_t\cdot {\bf A}_p){\bf B}_n -({\bf
A}_p\cdot {\bf B}_t){\bf V}_n]\cdot d{\bf S},
\end{aligned}
\end{equation}
where the magnetic field ${\bf B}$ and the velocity field ${\bf V}$ are observed quantities in the solar atmosphere, and the boundary value of the magnetic vector potential ${\bf A}_p$ of the reference field can be inferred from the vertical component of the magnetic field $B_n$. The subscripts $n$ and $t$ indicate the normal and transverse components, respectively. There, the first term in the right-hand side of equation (\ref{eq:heliflux}) is neglected, and the second one has been simplified from the observations of magnetic fields on the solar surface. 

The first term in the right-hand side of equation (\ref{eq:heliflux1}) provides the contribution from the horizontal motion of footpoints of the magnetic field at the solar surface, while the second one reflects the contribution from the vertical motion of magnetic flux at the surfaces of the integral.

Both equations (\ref{eq:heliflux3}) and (\ref{eq:heliflux1}) have been used to analyze the evolution of magnetic helicity from different perspectives. The former equation has been used in the study of the magnetic helicity with the dynamo process, and the latter on the injected flux of magnetic helicity from the enveloping surface, such as the solar surface. Agreement between the two methods can be taken as an indication of the validity of the analysis concerning the evolution of large-scale magnetic fields over the solar cycle.

According to the analysis of \cite{Demoulin03}, one finds that
\begin{equation}
\label{eq:heliflux2}
\frac{dH_{\rm m}}{dt}=-2\oiint_S({\bf U}\cdot {\bf A}_p){\bf B}_n\cdot d{\bf s},
\end{equation}
where
\begin{equation}
\label{eq:dember}
{\bf U}={\bf V}_t-\frac{V_n}{B_n}{\bf B}_t.%\nonumber
\end{equation}
This implies that the influence of the second term in equation (\ref{eq:heliflux1}) has been omitted as only the line of sight magnetograms have been used. It is also noticed that the three components of the velocity field in the photosphere can also be derived from the Differential Affine Velocity Estimator for Vector Magnetograms \citep[DAVE4VM;][]{Schuck08}. As one neglects the second term on the right-hand side of equation (\ref{eq:heliflux1}), there is an error of about 10\% in the calculation of the injected magnetic helicity as compared with the method DAVE4VM, as pointed out by \cite{Liu12} from their calculated results.  

Nevertheless, equation (\ref{eq:heliflux2}) is still important, as only the longitudinal component of the magnetic fields can be used. Notably, equation (\ref{eq:heliflux2}), combined with the local correlation tracking (LCT) method \citep{cha01} for calculating the horizontal velocity field of magnetic features, has been employed to analyze the large-scale injected magnetic helicity at the solar surface of cycle~23 using MDI magnetic synoptic charts \citep{YangSB12} and MDI 96 min full-disk magnetograms \citep{Zhang13}. Moreover, it reveals a similar tendency in the large-scale reversal of the sign distributions of magnetic helicity with solar cycles in the northern and southern hemispheres when compared with the calculation of the mean current helicity of active regions by \cite{zetal10}, and some comparable analysis by \cite{liu22}.

\section{Helicity injection in cycles 24 and 25}

In the subsequent analysis, we investigate the injection of magnetic helicity during Solar Cycle 24 and the early phase of Cycle 25, employing the methodologies proposed by \cite{YangSB12} and \cite{Zhang13}.
We also examine the distribution of magnetic helicity flux in the butterfly diagram,
aiming to deepen our understanding of the evolution of solar magnetic fields and the underlying physical mechanisms.
Using solar synoptic magnetic charts, we analyze the injected helicity during these cycles and its spatiotemporal distribution across time and latitude.

\subsection{Injected Magnetic helicity from individual active regions}

\begin{figure}
\begin{center}
	\includegraphics[width=80mm]{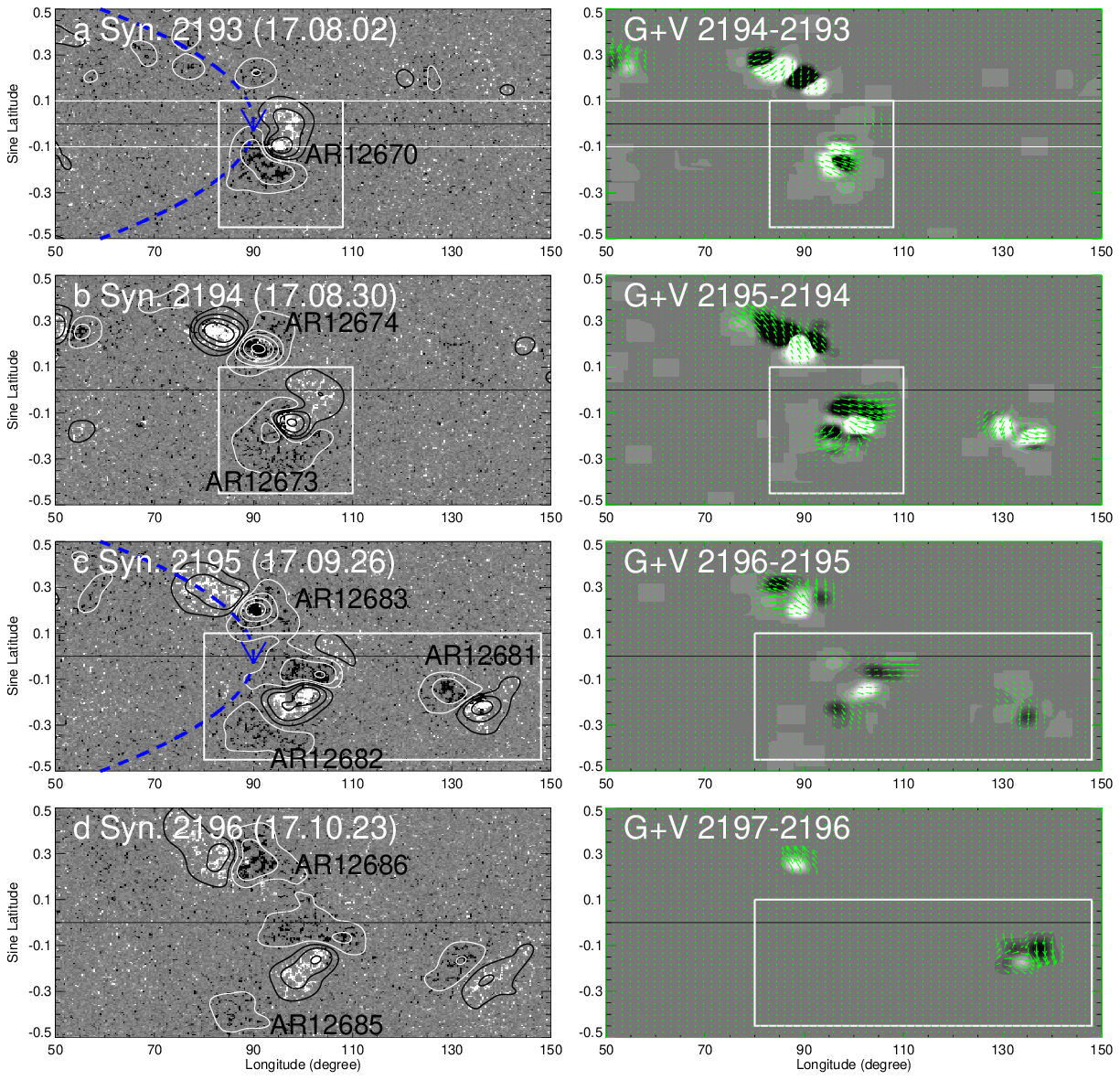}
 \end{center}
%\vspace{5mm}
    \caption{ Left: Evolution of large-scale magnetic fields of active region NOAA 12673 and the relationship with active regions NOAA 12670, 12682, and 12685 in the synoptic charts of different Carrington rotations 2193--2196. The blue dashed lines with arrows mark the regular direction of magnetic fields in the convection zone according to the Hale polarity law of magnetic fields. The contour levels (black/white) of the smoothed magnetic fields are $\pm$5, 20, 50, 100, and 200 G. Right: injected rate $G=-({\bf V}_t\cdot {\bf A}_p)B_n$ (white (black) shows the positive (negative) sign), and the horizontal velocity (arrows) inferred by LCT.  }
    \label{Fig:syncharts}
\end{figure}

\begin{figure}
\begin{center}
	\includegraphics[width=80mm]{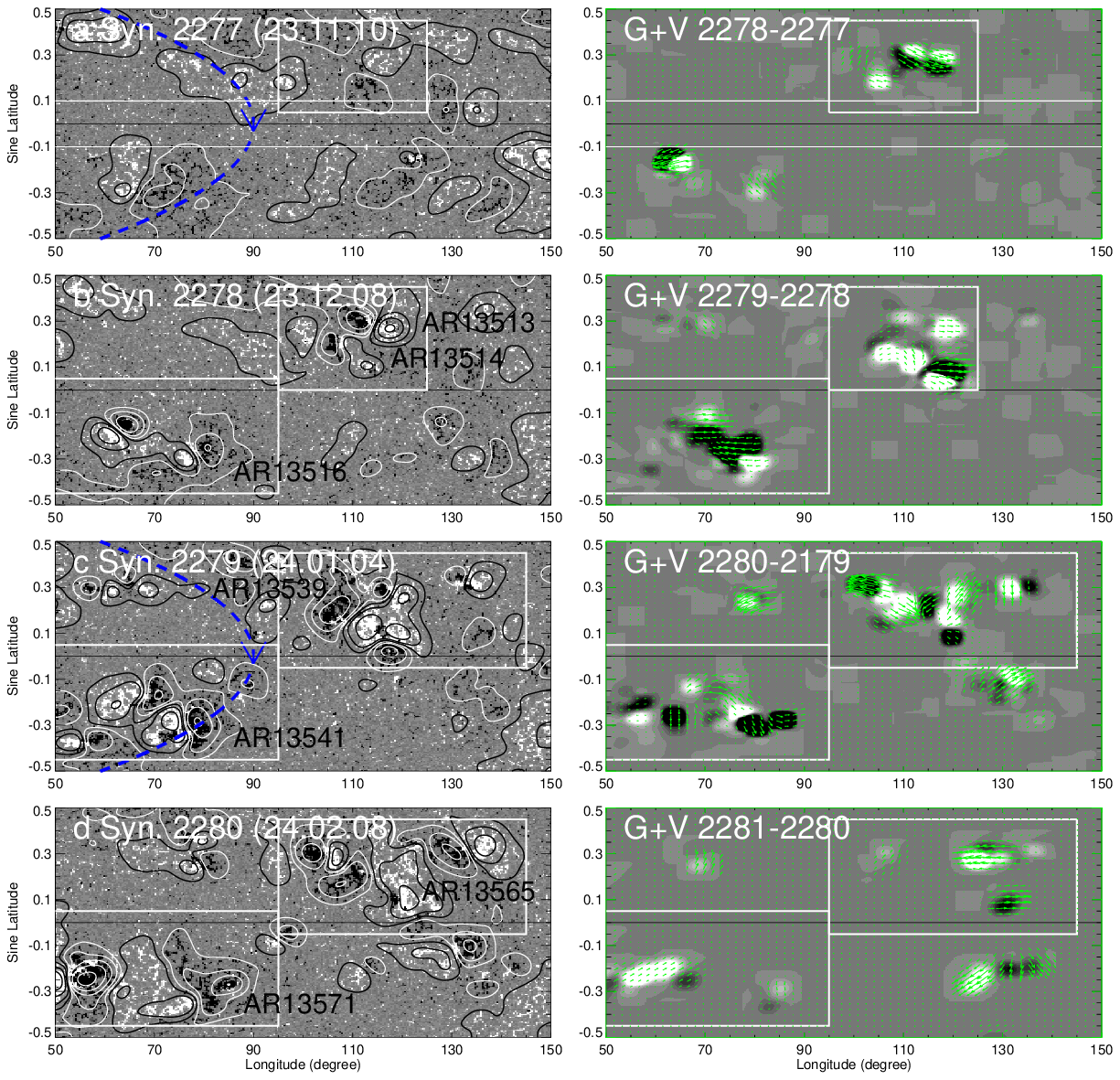}
 \end{center}
\vspace{5mm}
    \caption{ Left: Evolution of large-scale magnetic fields of active region NOAA 13513-13514 and 13516, and the following active regions at similar active longitudes in the synoptic charts of different Carrington rotations 2277--2280. Right: The same as Fig.~\ref{Fig:syncharts}. }
    \label{Fig:synchartsa}
\end{figure}

The statistical analysis of the emergence of magnetic flux with twisted magnetic loops from the subsurface was proposed by \cite{Longcope98}. Although it is favorable to compare with observations in both mean value and statistical dispersion, it is also of paramount importance to track the evolution process of magnetic flux with helicity in individual active regions. This is because the unique characteristics and dynamic behaviors within each active region can offer crucial insights into the complex physical mechanisms underlying solar eruptions and magnetic field restructuring.

Figure~\ref{Fig:syncharts} shows a series of synoptic charts of the magnetic field at different CRs and the corresponding time difference. It provides an opportunity to analyze the large-scale magnetic fields for the detection of a possible evolution of magnetic fields from the subsurface. This is a sequence of solar active regions NOAA 12670, 12673, 12682, and 12685. It is found that the leading and following polarities of the large-scale magnetic bipoles change in time before and after the emergence of active region NOAA 12673 \citep[cf.][]{Yang17,Sun17,Romano19}. This provides evidence for a reversed variation of the observed large-scale magnetic bipole of active regions due to the emergence of negative magnetic helicity flux from the deeper solar convection zone. This is similar to the highly sheared magnetic field that occurred in the active region NOAA12681 in the southern hemisphere with the injection of magnetic helicity.

Figure~\ref{Fig:syncharts} also shows the injected rate $G\!=\!-({\bf V}_t\cdot {\bf A}_p)B_n$ calculated from the synoptic magnetic charts and the corresponding horizontal velocity ${\bf V}_t$ by LCT after some smoothing. The horizontal velocity arrows, resulting from the evolution of large-scale magnetic fields on the solar surface, are inferred from LCT. The consistency in the direction of the velocity field with the evolving direction of the large-scale magnetic fields can normally be found, for example, in the active regions NOAA 12670, 12673, and 12682. It is also noticed that the contribution of the helicity in the quiet Sun is negligible.

For comparison, Fig.~\ref{Fig:synchartsa} presents another sample, the synoptic magnetic charts corresponding to Carrington rotations 2277--2280,
in the early phase of cycle 25, which includes active regions NOAA13516, 13539, 13541, 13565, 13571 et al. that occurred at different solar rotations.

It is found that a major contribution of the magnetic helicity $G\!=\!-({\bf V}_t\cdot {\bf A}_p)B_n$ comes from the active regions, even if this only provides the variation of magnetic fields on the synoptic scale, where the relatively small temporal scale ones have been ignored. However, compared with all solar cycles, it can also be presented as the contribution from the fluctuating magnetic fields. From the evolution of the large-scale magnetic fields in Fig.~\ref{Fig:syncharts} and \ref {Fig:synchartsa}, we suggest that these reflect the local exchange between the poloidal and toroidal fields emerging in the subsurface.

\begin{figure}
\begin{center}
%\hspace{25mm}	
\includegraphics[width=70mm]{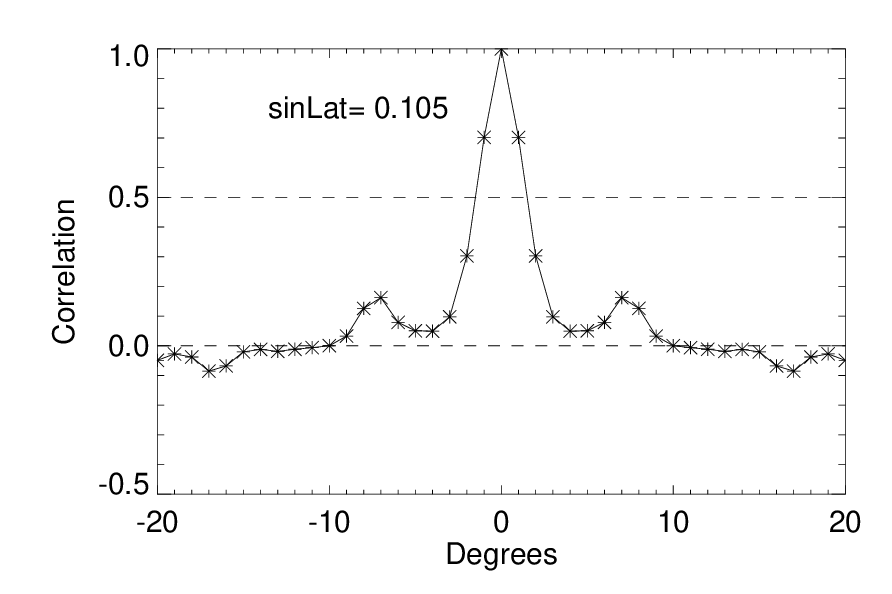}
\includegraphics[width=70mm]{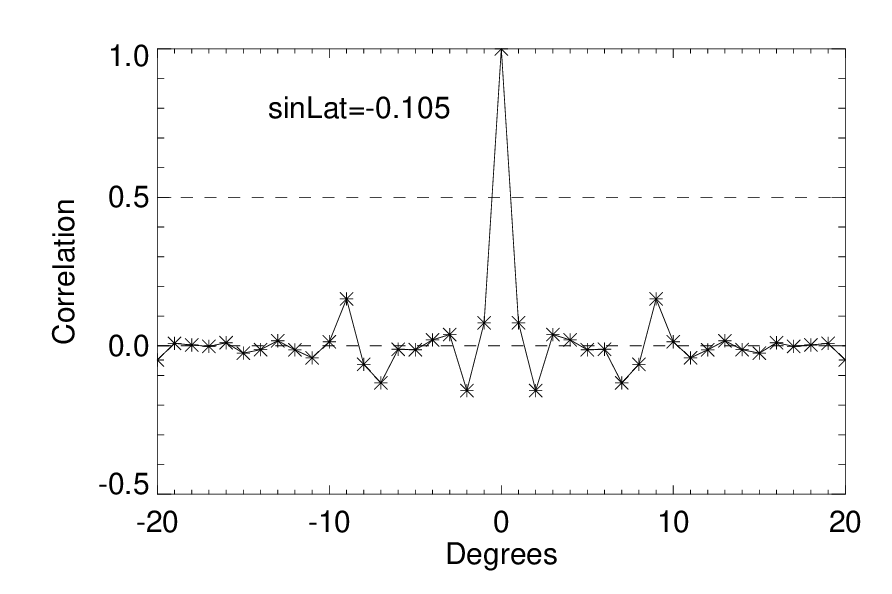}
 \end{center}
\vspace{0.05mm}
\caption{Autocorrelation function as a function of latitude along the time direction of  CR2193
(labeled by two straight white horizontal solid lines in Fig.~\ref{Fig:syncharts}a.)
Top: relative to the active region, the low line, and bottom: the quiet Sun, the top line in Fig.~\ref{Fig:syncharts}a.
}\label{fig:autocorrelation}
\end{figure}

In Figs.~\ref{Fig:syncharts} and \ref {Fig:synchartsa}, it is seen that the contribution of the horizontal velocity with the injected magnetic helicity occurs in local areas of active regions. This occupies only a relatively small amount in the magnetic charts due to the smaller contribution from the quiet Sun.

Since the horizontal velocity is obtained by using the LCT method, some magnetic helicity signals may be lost. Here, we use the autocorrelation function to evaluate the loss. As an example, Fig.~\ref{fig:autocorrelation} shows the auto-correlation function for different latitudes along the time direction of CR~2193 in Fig.~\ref{Fig:syncharts}a. It is found that the auto-correlation function is much sharper in the region without sunspots. While using the LCT method, this may cause a low correlation and underestimate the movements at small scales. Thus, we obtain a relatively small amplitude of magnetic helicity fluxes compared with previous results, such as those of \cite{YangSB12} and \cite{Zhang13}. However, this could still reflect helicity flux characteristics at the specific spatial resolution and temporal cadence. This is also consistent with the local contribution of injected helicity from the large-scale field distributed in Fig.~\ref{Fig:syncharts}a.

Using the synoptic chart data processing methodology described above, we will analyze magnetic helicity injection as a function of solar cycle phases in the subsequent analysis.

\subsection{Distribution of injected magnetic helicity in solar cycle 24 and the early phase of cycle 25}

\begin{figure*}
 \vspace{7mm}
  \begin{center}
 	\includegraphics[width=150mm]{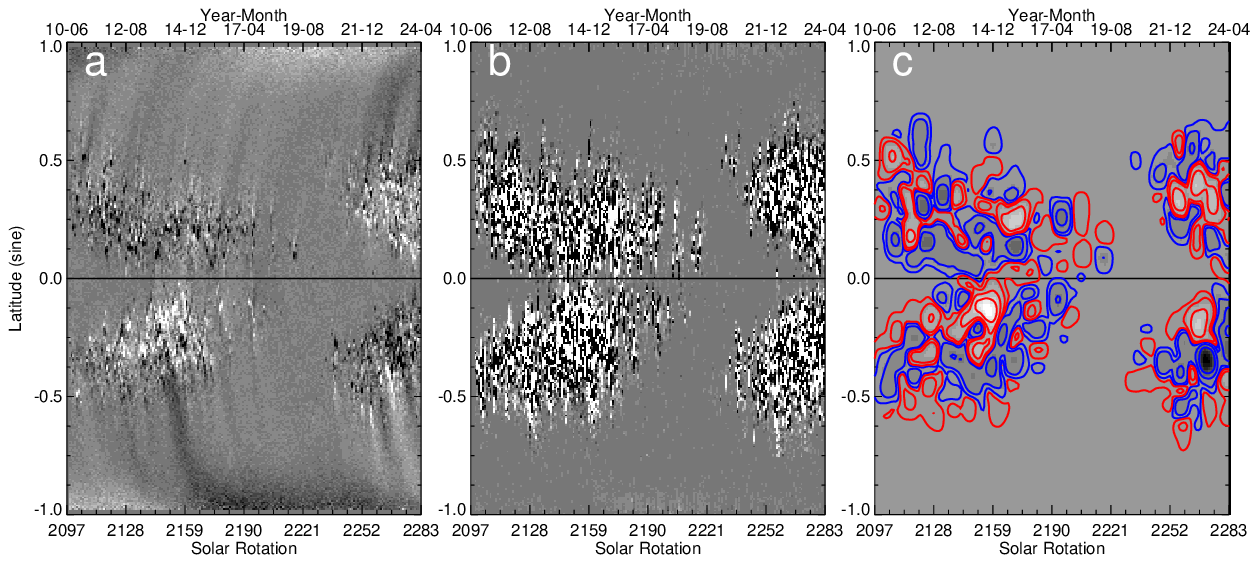}	
 \end{center}
 \vspace{5mm}
    \caption{ 
 a) Butterfly diagram of magnetic fields in solar cycles 24 and part of 25. White (black) represents positive (negative) polarity. b) Distribution of the injection rate of magnetic helicity in both hemispheres. White (black) indicates positive (negative) values. c) Injection rate of magnetic helicity in both hemispheres after smoothing. Red (blue) contours denote the relative injection rates of 1, 5, 20, 50 $\times 10^{32}$(Mx)$^2/s$ with positive (negative) values.   
    }
    \label{fig:butterfly1}
\end{figure*}

It is generally believed that the magnetic field and the corresponding helicity at the solar surface are important parameters.
These parameters help us understand the possible effects on the solar dynamo process in the solar convection zone, where the $\alpha$ effect is one of the key parameters in this process \citep[e.g.][]{par55,Steenbeck66,Zeldovich83}.
The magnetic and current helicities are important ingredients in the $\alpha$ effect; see \cite{bra-sub:05} for a review.

The collective effect of the active regions on the contribution of magnetic and current helicities over the solar cycle has been studied from observations
\citep[e.g.,][]{zetal10,Yangx12,ZhangBS2016} and also with solar dynamo models \citep[see, e.g.,][]{ketal03,Choudhuri04,Xu09,pietal13mn}.

\begin{figure}
 \vspace{7mm}
\begin{center}
	\includegraphics[width=70mm]{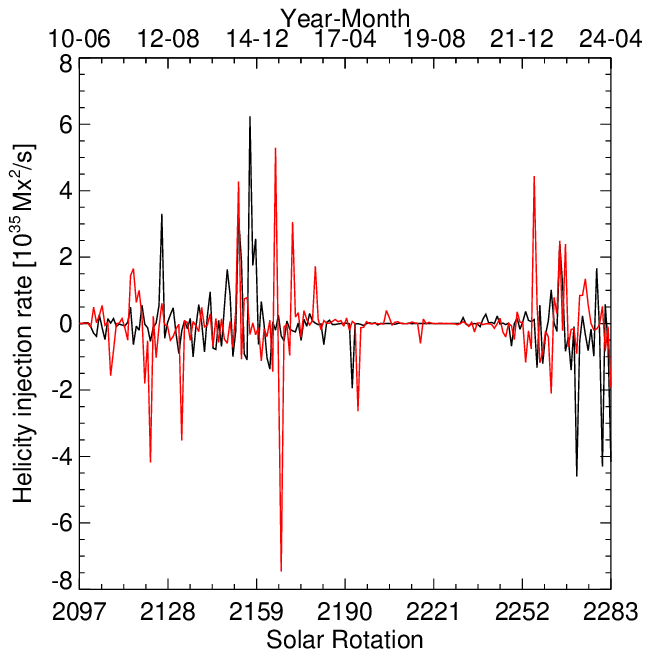}
 \end{center}
 \vspace{5mm}
    \caption{
The injection rate of magnetic helicity in the northern (red) and southern (black) hemispheres as a function of latitude during solar cycle 24 and the early phase of cycle 25.}
    \label{fig:butterfly3}
\end{figure}

\begin{figure*}
\vspace{7mm}
\begin{center}
	\includegraphics[width=150mm]{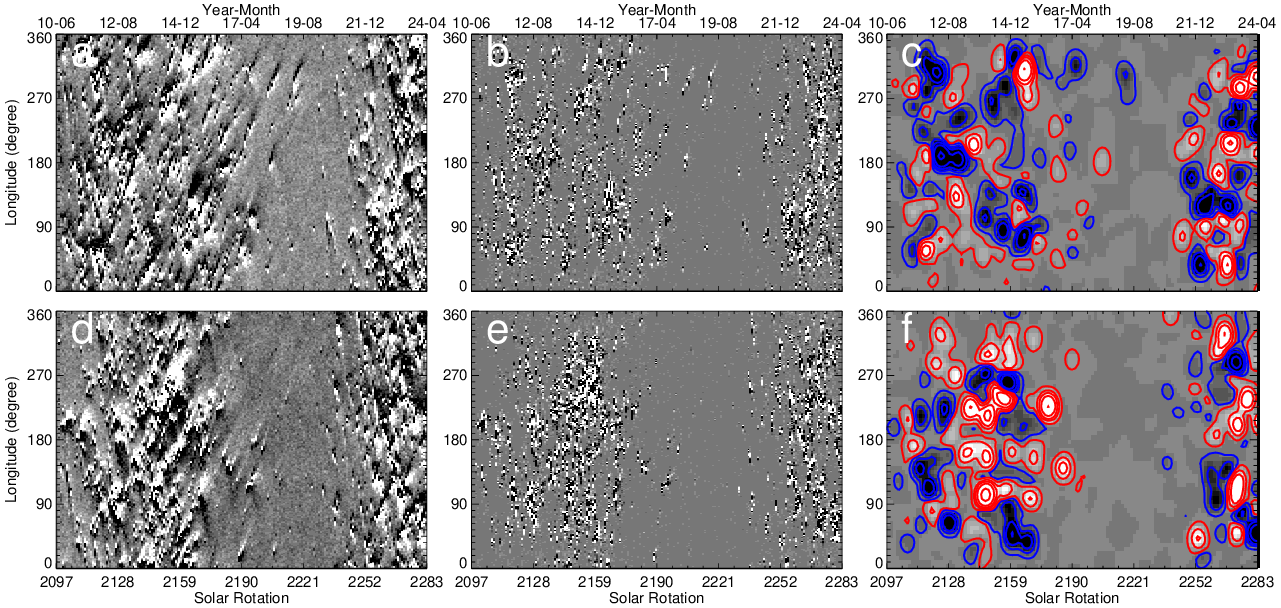}
 \end{center}
\vspace{5mm}
\caption{
Left: Mean magnetic field as a function of longitude in the northern (a) and southern (d) hemispheres during solar cycle 24 and the early phase of 25.
Middle: The corresponding injection rate of magnetic helicity in both hemispheres (b and e).
Right: Injection rate of magnetic helicity in both hemispheres (c and f) after smoothing.
Red (blue) contours represent the injection rates of 1, 5, 10, and $20\times 10^{32}\,{\rm Mx}^2\,{\rm s}^{-1}$ with positive (negative) signs.
}
\label{fig:longiheli}
\end{figure*}

In this study, 184 HMI magnetic synoptic charts from June 2, 2010, to April 22, 2024 have been used in the calculation to obtain the longitudinal butterfly diagrams in Fig.~\ref{fig:butterfly1}a. These charts cover most of the large-scale magnetic fields during solar cycle 24 and the early stage of cycle 25. It is noted that most of the mean negative fields migrate to the equator at the low latitude in the northern hemisphere in cycle 24, and that of the positive polarity in the southern hemisphere.
In contrast, cycle 25 exhibits the opposite polarity migration pattern.
The dominant single polarity at low latitudes in the solar synoptic magnetogram indicates that the opposite-polarity magnetic field there has been partially weakened,
and some of it migrated to the polar regions \citep{Wangy89} along with the meridional flow in the solar convection zone \citep{Zhao13,Gizon21}.
This meridional flow acts as a conveyor belt, transporting magnetic flux across the solar surface and redistributing it, which plays a crucial role in the long-term evolution of the solar magnetic field.

To analyze the transfer of the magnetic helicity contributed from the magnetic fields,
Fig.~\ref{fig:butterfly1}b shows the butterfly diagram of the mean injected magnetic helicity inferred from a series of HMI synoptic charts employing equation (\ref{eq:heliflux2}).
It shows the mean values of injected helicity with latitude at various solar Carrington rotation cycles.
This result reflects the contribution of large temporal and spatial scale magnetic fields to the magnetic helicity in the magnetic synoptic charts.
This means that the results reflected in Figs~\ref{fig:butterfly1}a and \ref{fig:butterfly1}b show similar temporal and spatial scales.

It was pointed out by \cite{ZhangBS2016} that around the solar maximum, the magnetic energy and helicity spectra are steeper, emphasizing the large-scale field.
This tendency can also be found in Fig.~\ref{fig:butterfly1}, where the large-scale structures are the result of smaller elements of magnetic field and helicity of the same sign
during solar maximum.

From Fig.~\ref{fig:butterfly1}, it is found that the injected magnetic helicity at the solar surface shows a trace to the equator with the migration of the large-scale magnetic field and also shows a tendency toward the northern and southern poles. Still, it does not find a significant signature of the helicity transport to the polar regions. It reflects that most of the helicity has been erupted with the evolution of magnetic fields of active regions, thereby making it difficult to reach the solar polar regions. Figure~\ref{fig:butterfly1}b shows the injected magnetic helicity with the variation of magnetic features corresponding to temporal and spatial scales calculated from a series of synoptic magnetic charts.

Figure~\ref {fig:butterfly1}c presents the smoothed injected helicity during solar cycle 24 and the early phase of cycle 25 to illustrate the mean contribution of magnetic helicity. 
It is easy to observe that relatively strong large-scale positive injected helicity patterns appear near Carrington rotations (CRs) 2159 and 2167 in the southern hemisphere, as well as near CRs 2170 and within the range of 2260--2280 in the northern hemisphere. Conversely, negative injected helicity patterns are detected between CRs 2130 and 2167 in the northern hemisphere, and around CR 2267 in the southern hemisphere.
It is also worth emphasizing that the accuracy of calculating magnetic helicity in high latitudes is low due to the influence of projective effects on the observed magnetograms and the corresponding synoptic charts.

Figure~\ref{fig:butterfly3} shows the mean injected rate of magnetic helicity in the northern and southern hemispheres during solar cycle 24
based on the calculation of a series of HMI synoptic magnetic charts.
The mean values in the northern and southern hemispheres are $-83.7\,{\rm Mx}^2\,{\rm s}^{-1}$ and $66.6\,{\rm Mx}^2\,{\rm s}^{-1}$, respectively.
These mean values, combined with the visual representation in Fig.~\ref {fig:butterfly3},
provide a rough estimation of the distribution of the injected magnetic helicity in the solar atmosphere.
The total flux of injected magnetic helicity in this calculation is on the order of $10^{44}\rm Mx^2$ in the whole solar cycle 24,
while it is lower than $5\times10^{46}\,{\rm Mx}^2$, as estimated by \cite{Zhang13}, because this calculation is based on the magnetic synoptic charts,
some short temporal and spatial magnetic flux and its contribution for the helicity flux have been ignored.

The difference in the estimated total magnetic helicity from the solar surface between the results of \cite {Zhang13}, which used MDI 96-minute full-disk magnetograms, and our calculation based on magnetic synoptic charts is on the order of $5\times10^2$. The difference in the temporal intervals of the magnetic field data series used for calculating the injected helicity between our study and that of \cite{Zhang13} is on the order of $4\times10^2$. These significant differences in the estimated values and data intervals imply important implications for the calculation of injected helicity. This means that a certain amount of injected helicity has been lost in our calculation. This also indicates that the results of \cite{Zhang13} provide a lower estimate of the injected helicity from the solar surface during the solar cycles. This is because they only neglected the contribution from short time scales of less than 96 minutes.

Figure~\ref{fig:longiheli} shows the distribution of the mean magnetic field with longitude in the northern and southern hemispheres in solar cycle 24 and the early phase of cycle 25, and the corresponding injected rate of magnetic helicity inferred from the magnetic synoptic charts, which are contributed by the magnetic fields of a series of active regions. In Figs \ref{fig:longiheli}a and d, it is easy to see the slanted arrangement of a series of magnetic field structures like scratches in both hemispheres caused by the differential rotation of the Sun. One can also see a large-scale pattern of the injected helicity after the data are smoothed in Fig.~\ref{fig:longiheli}c and \ref{fig:longiheli}f. It is consistent with the idea that the dominant sign of the injected helicity shows a negative sign in the northern hemisphere and a positive one in the southern hemisphere in solar cycle 24. It is also true of Fig.~\ref{fig:butterfly1}c.
The relatively strong, large-scale negative injected helicity patterns occur near CR~2128 in the northern hemisphere and positive ones near CR~2135 in the southern hemisphere.
This is roughly consistent with the result in Fig.~\ref{fig:butterfly1}c. However, when it comes to the early phase of cycle 25, the sign rule of the injected magnetic helicity is not as obvious in Fig.~\ref {fig:longiheli} as it was in cycle 24.

By comparing with Fig.~\ref{fig:butterfly1}a, it is found that the dominant sign of the magnetic fields reverses between cycle 24 and 25 at the low latitudes. It is consistent with Hale's polarity law. Therefore, it is found that the dominant contribution to the injected magnetic helicity likely comes from the relatively strong magnetic fields of the leading polarity of active regions interacting with large-scale velocities induced by solar differential rotation, as some of the weak fields have been discarded or drowned in the synoptic charts. The differential rotation with large-scale magnetic fields can be found in Fig.~\ref{fig:longiheli}a.  It can also be compared to the results from the full-disk magnetograms by \cite{Zhang13}. In their calculation that the positive and negative fields are nearly balanced, and the sign rule of injected magnetic helicity is observed. Thus, discrepancies exist in the calculated magnetic helicity results for the solar northern and southern hemispheres when using different methods and data. Nevertheless, these differences are self-consistent and complementary.

\begin{figure}
\begin{center}
	\includegraphics[width=80mm]{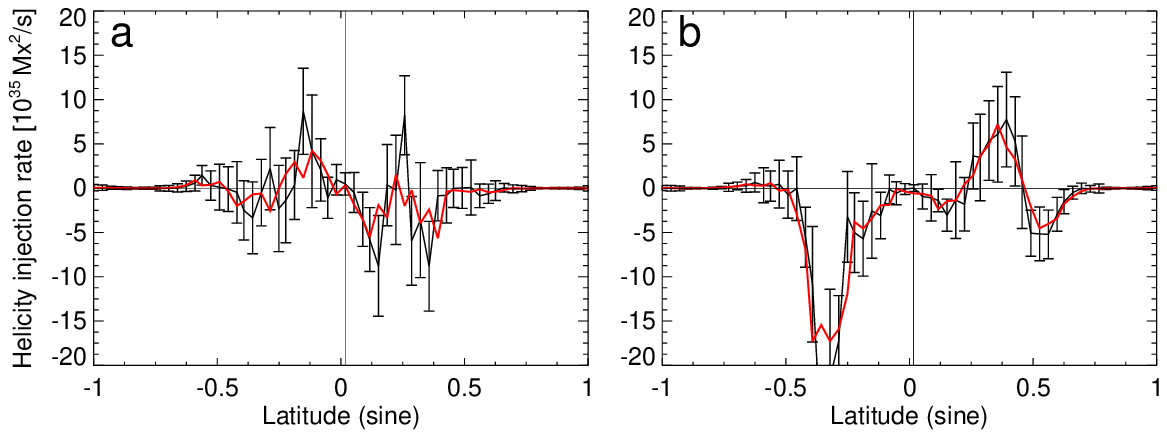}
 \end{center}
% \vspace{5mm}
    \caption{ The mean injection rate of magnetic helicity as a function of latitude in solar cycle 24 (left) and the early phase of solar cycle 25 (right). The error bars represent only the relative deviation of the mean values. The red line indicates the distribution of injected magnetic helicity after smoothing.
}
    \label{fig:butterfly4}
\end{figure}

\begin{figure}
\begin{center}
	\includegraphics[width=80mm]{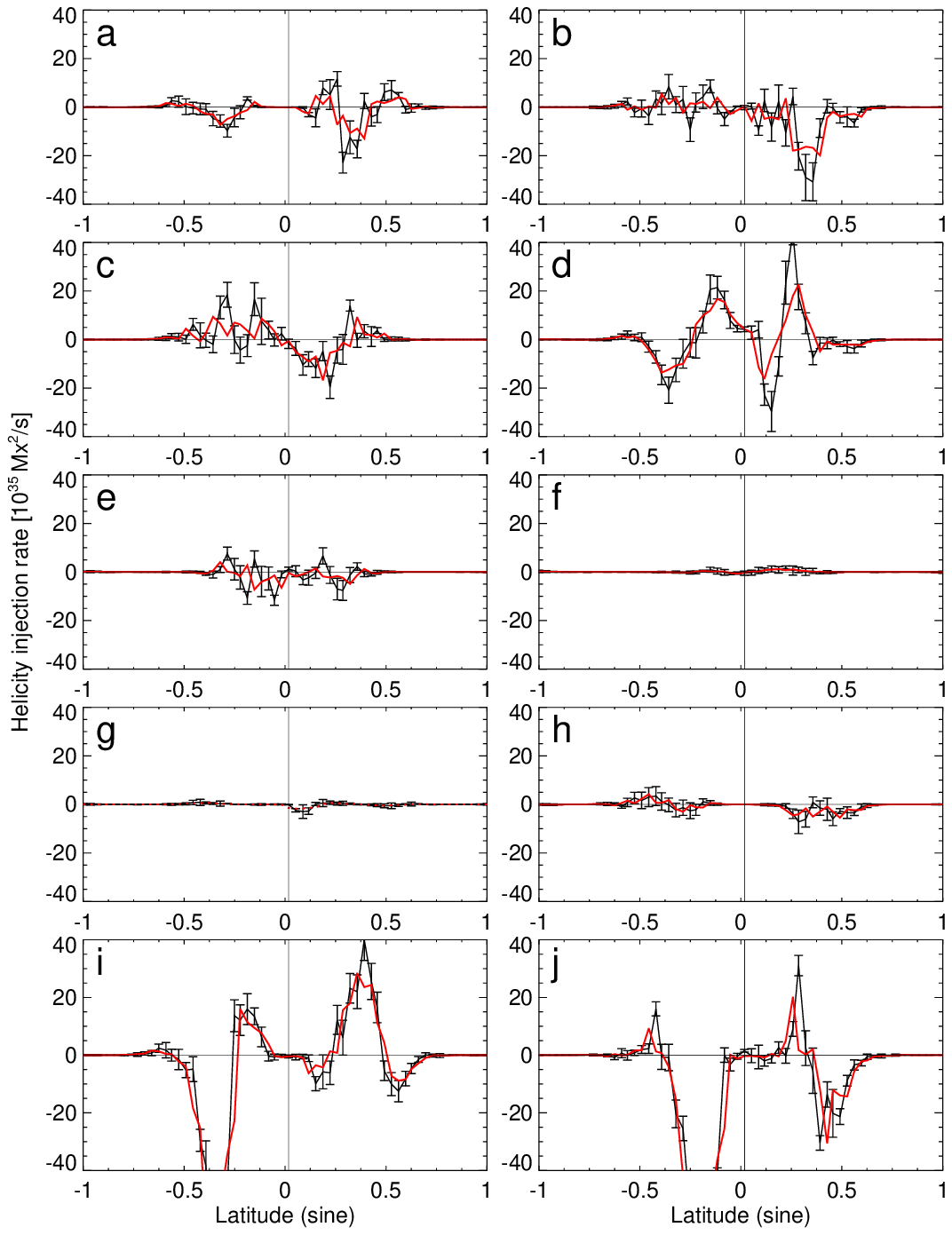}
 \end{center}
 %\vspace{5mm}
\caption{The mean injection rate of magnetic helicity as a function of latitude in solar cycle 24 and the early phase of cycle 25,
after averaging over groups of 20 solar synoptic charts from panels a to i, and panel j is used for the remaining few charts.
The error bars only represent the relative deviation of the mean values. The red line indicates the distribution of injected magnetic helicity after smoothing.}
    \label{fig:butterfly4a}
\end{figure}

Figure~\ref{fig:butterfly4} shows the mean injected rate of magnetic helicity with latitude.
It is found that the dominant sign is negative in the northern hemisphere and positive in the southern hemisphere in solar cycle 24
in Fig.~\ref{fig:butterfly4}a.
The reversal relative to the helicity sign rule is shown in the early phase of solar cycle 25 in Fig.~\ref{fig:butterfly4}b.

In order to analyze the evolution of injected magnetic helicity flux with latitude, the injected magnetic helicity flux, which was calculated from solar synoptic charts,
was divided into 10 segments, as shown in Fig.~\ref {fig:butterfly4a}.
It can be found that panels a-e belong to solar cycle 24, while panels h--j of this figure belong to the rising phase of solar cycle 25.
Based on this categorization, it is found that the significant helicity flux amplitudes occur at solar rotation 2159--2179 in solar cycle 24,
and even larger amplitudes occur at 2259--2283 in solar cycle 25.
These results are consistent with the fact that solar cycle 25 is a stronger cycle than cycle 24.

\begin{figure}
 \vspace{8mm}
 \begin{center}
 	\includegraphics[width=60mm]{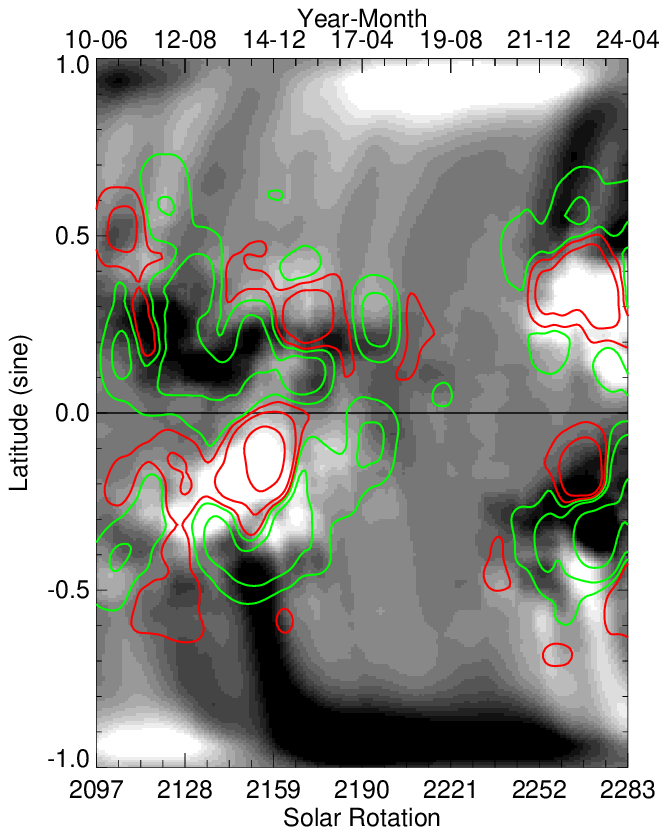}	
 \end{center}
 \vspace{3mm}
    \caption{ 
The smoothed butterfly diagram of the line-of-sight magnetic field in solar cycles 24 and part of 25.
White (black) represents positive (negative) polarity overlaid with the injection rate of magnetic helicity in both hemispheres.
Red (green) contours denote the relative injection rates of 1, 5, 20 $\times 10^{32}\,{\rm Mx}^2\,{\rm s}^{-1}$ with positive (negative) values.
}\label{fig:butterfly-smooth}
\end{figure}

Figure \ref{fig:butterfly-smooth} depicts the large-scale magnetic fields and injective magnetic helicity during solar cycle 24 and the early phase of cycle 25. The data have been smoothed to emphasize low-frequency (large-scale) magnetic features and investigate their relationship with corresponding large-scale injected magnetic helicity.  Notably,  at low latitudes in the butterfly diagram, positive magnetic fields exhibit a statistical correlation with positive injected helicity, whereas negative fields show a similar correlation with negative injected helicity. We also observe that relatively weak large-scale injected magnetic helicity with opposite signs tends to appear in the edge regions of the solar magnetic butterfly diagrams.   

When we note that the horizontal velocity of higher latitudes decelerates relative to lower latitudes due to solar differential rotation, and observe in detail that the magnetic fields in the northern hemisphere of solar cycle 24 predominantly show negative polarity in the magnetic butterfly diagrams in Fig.~\ref{fig:butterfly-smooth}, the result of large-scale negative magnetic helicity transport is naturally obtained. Similarly, positive magnetic helicity transport in the southern hemisphere of cycle 24 can be deduced.
In contrast to the dominant signs of magnetic helicity in the northern and southern hemispheres in cycle 24,
which are opposite to those in the early phase of cycle 25, it is found that the statistical sign of magnetic helicity transport in both hemispheres during the early phase of cycle 25 is reversed relative to the general hemispheric sign rule of helicity.
It is worth noting that similar reversed current helicity patterns have also been presented at the beginning of earlier solar cycles 22-23 (e.g. \citealt{bao00}, see also \citealt{zetal10}). 
The disturbances caused by the injection of reversed magnetic helicity in strong active regions constitute factors that cannot be ignored, as has been noted by observations \citep[e.g.][]{liu06,Zhangy08,Zhang13}.

It should be noted that the solar magnetic butterfly diagram shows that the large-scale magnetic fluxes in the northern and southern hemispheres exhibit opposite polarities.
In principle, the magnetic fluxes from solar active regions or quiet regions should be statistically balanced.
As part of the magnetic field migrates to high latitudes or the polar regions of the Sun due to the effect of meridional circulation
and the following magnetic fields of active regions or weak magnetic fields in parts of the quiet regions get attenuated in the process,
a statistical imbalance of the large-scale magnetic fields will occur, and the calculated large-scale injected magnetic helicity may deviate from the equilibrium state.

When analyzing the large-scale magnetic helicity transport in the Sun using synoptic magnetic charts,
seemingly contradictory results may emerge.
It should be pointed out, however, that when analyzing the relationship between the solar magnetic field and magnetic helicity from multiple perspectives,
we should also take into account that the results were obtained using different methods.
In particular, the magnetic helicity that is transported from the solar interior should in principle be statistically balanced between the northern and southern hemispheres,
but there are always fluctuations, which can be found both from the perspective of current helicity in solar active regions \citep{Seehafer90,Pevtsov95,BZ98,Hagino05,zetal10}
and from the injection of magnetic helicity from the full disk magnetograms \citep{YangSB12,Zhang13}; see also \cite{BergerRuzmaikin00}.
Our observational results capture the injected helicity from a subset of magnetic fields from synoptic magnetic charts.
This should be kept in mind when comparing with the results from other studies, which may reflect the magnetic helicity transport from different perspectives.
These findings are therefore not contradictory, but may well be mutually consistent.

\section{Discussion and conclusions}

In this paper, the collective effect of injected magnetic helicity during solar cycles and individual samples of magnetic helicity with the evolution of the magnetic field in active regions have been presented. To analyze the helicity formation in the solar subsurface, the collective effect of the injected magnetic helicity with the magnetic fields of solar cycle 24 and the early phase of cycle 25 has been presented based on the analysis of a series of magnetic synoptic charts. The sign distribution of injected magnetic helicity shows the dispersed form contributed from individual local regions (such as active regions) with solar cycles, and it is hard to find more helicity in the polar regions, due to the eruption of the magnetic field from the sub-atmosphere into the interplanetary space.

It is found in our calculations that the mean values of the injected magnetic helicity flux are consistent with the hemispheric sign rule of current helicity of solar active regions,
i.e.\ the negative (positive) sign tends to be in the northern (southern) hemisphere in solar active cycle 24.
The reversal pattern relative to the sign rule occurs in the early phase of cycle 25.
Furthermore, according to Hale's law, the polarities of active regions statistically reverse between solar cycles 24 and 25.
Therefore, the dominant contribution to the injected magnetic helicity likely comes from the relatively strong magnetic fields of the leading polarity of active regions
interacting with the local velocity fields around them, wrapped by large-scale velocity fields due to the solar differential rotation, 
as part of the magnetic fields is advected to high latitudes
or polar regions of the Sun due to the effect of meridional circulation.
A similar effect is found for the following magnetic fields of active regions or weak magnetic fields in parts of quiet regions that are attenuated in the data processing.
Overall, further multi-faceted analyses are required to fully understand the injection of magnetic helicity on the solar surface and its contribution from different sources.

For presenting the evolution of magnetic helicity with solar cycles, a morphological analysis of the helicity patterns with butterfly diagrams has been proposed,
which is composed of different scale fluctuating components, such as those characterized by solar active regions.
This probably reflects the complex process of twisted magnetic fields inside the convection zone
due to the interaction of the Coriolis force and meridional circulation on the formation of magnetic field lines.

We have analyzed the evolution of the magnetic helicity in active regions in different solar Carrington rotation cycles.
The injected magnetic helicity occurs on the solar surface with the emergence of magnetic flux in active regions.
This probably reflects the emergence of twisted new magnetic flux formed in the deep convection zone,
and some of the flux interacts with the existing magnetic field of the active regions on the solar surface.
This process is associated with the injection of magnetic helicity.
Furthermore, it is probably accompanied by the exchange of different-scale poloidal and toroidal fields.

Although the injected magnetic helicity flux has been calculated based on synoptic magnetic charts at the relevant temporal and spatial scales of magnetic fields, a detailed analysis of how the magnetic helicity evolution contributes to the solar cycle requires considering more aspects observed in full-disk vector magnetograms.
However, when the observed solar magnetic fields of different temporal and spatial scales are used in the analysis, the statistical results may show some differences, though we may hope that the trends still reflect the fundamental characteristics of magnetic fields from different perspectives.

\section*{Acknowledgements}

The authors would like to thank %the referee for his/her comments for improving the manuscript, and
the staff at the \textit{Solar Dynamics Observatory} (SDO) for providing the observational data. The authors would like to thank Dr.\ V.\ Pipin for useful discussions. This study is supported by grants from the National Natural Science Foundation (NNSF) of China under the project grants 11673033, 11427803, 11427901, 11703042, 11911530089, 12073040, 120730041, 12473052, and other grants at Huairou Solar Observing Station, National Astronomical Observatories, Chinese Academy
of Sciences. This work was supported in part through the Swedish Research Council, grant 2019-04234.
K.K. would like to acknowledge support under the PIFI program of the Chinese Academy of Sciences and thank Huairou Solar Observing Station, National Astronomical Observatories of China, for hospitality.

The original data source: http://jsoc1.stanford.edu/

%\newpage

\end{document}